\documentclass[twocolumn,showpacs,preprintnumbers,amsmath,amssymb,superscriptaddress]{revtex4}

\usepackage{graphicx}
\begin{document}
\title{Madelung potentials and covalency effect 
in strained La$_{1-x}$Sr$_x$MnO$_3$ thin films 
studied by core-level photoemission spectroscopy}

\author{H.~Wadati}
\email{wadati@phas.ubc.ca}
\homepage{http://www.geocities.jp/qxbqd097/index2.htm}
\affiliation{Department of Physics and Astronomy, University of British
Columbia, Vancouver, British Columbia V6T-1Z1, Canada}

\author{A.~Maniwa}
\affiliation{Department of Applied Chemistry, University of Tokyo, 
Bunkyo-ku, Tokyo 113-8656, Japan}

\author{A.~Chikamatsu}
\affiliation{Department of Applied Chemistry, University of Tokyo, 
Bunkyo-ku, Tokyo 113-8656, Japan}

\author{H.~Kumigashira}
\affiliation{Department of Applied Chemistry, University of Tokyo, 
Bunkyo-ku, Tokyo 113-8656, Japan}
\affiliation{Core Research for Evolutional Science and Technology of Japan Science 
and Technology Agency, Chiyoda-ku, Tokyo 102-0075, Japan}
\affiliation{Synchrotron Radiation Research Organization, 
University of Tokyo, Bunkyo-ku, Tokyo 113-8656, Japan}

\author{M.~Oshima}
\affiliation{Department of Applied Chemistry, University of Tokyo, 
Bunkyo-ku, Tokyo 113-8656, Japan}
\affiliation{Core Research for Evolutional Science and Technology of Japan Science 
and Technology Agency, Chiyoda-ku, Tokyo 102-0075, Japan}
\affiliation{Synchrotron Radiation Research Organization, 
University of Tokyo, Bunkyo-ku, Tokyo 113-8656, Japan}

\author{T.~Mizokawa}
\affiliation{Department of Complexity Science and Engineering, 
University of Tokyo, Kashiwa, Chiba 277-8561, Japan}

\author{A.~Fujimori}
\affiliation{Department of Physics, University of Tokyo, 
Bunkyo-ku, Tokyo 113-0033, Japan}

\author{G.~A.~Sawatzky}
\affiliation{Department of Physics and Astronomy, University of British
Columbia, Vancouver, British Columbia V6T-1Z1, Canada}

\date{\today}
\begin{abstract}
We have investigated the shifts 
of the core-level 
photoemission spectra of La$_{0.6}$Sr$_{0.4}$MnO$_3$ thin films 
grown on three kinds of substrates, SrTiO$_3$, 
(LaAlO$_3$)$_{0.3}$-(SrAl$_{0.5}$Ta$_{0.5}$O$_3$)$_{0.7}$, 
and LaAlO$_3$. 
The experimental shifts of the La $4d$ and Sr $3d$ core levels are 
almost the same as the calculation, which we attribute to 
the absence of covalency effects on the Madelung potentials 
at these atomic sites 
due to the nearly ionic character of these atoms. 
On the other hand, 
the experimental shifts of the O $1s$ and Mn $2p$ core levels are 
negligibly small, in disagreement 
with the calculation. We consider that 
this is due to 
the strong covalent character of the Mn-O bonds. 
\end{abstract}
\maketitle
\section{Introduction}
The binding energy of a core level observed by photoemission
spectroscopy is determined by many factors. 
The shift $\Delta E$ of the energy of the core level 
measured relative to the chemical potential $\mu$, 
when the band filling is varied and/or the crystal 
structure is changed, is given by \cite{Hufner-book} 
\begin{equation}
\Delta E = - \Delta\mu + K\Delta Q - \Delta V_M + \Delta E_R, 
\end{equation}
where $\Delta\mu$ is the change in the chemical potential $\mu$, 
$\Delta Q$ is the change in the number of valence electrons 
on the atom, $K$ is a constant, $\Delta V_M$ is the change 
in the Madelung potential $V_M$ and $\Delta E_R$ is the change 
in the extra-atomic relaxation energy of the core-hole state. 
When the band filling is varied, both $\mu$ and $V_M$ should change, 
but the effect of $\Delta V_M$ has not been observed 
so far in experiments on transition-metal oxides. 
This has been most clearly demonstrated for 
La$_{2-x}$Sr$_x$CuO$_4$ \cite{Ino, pote}, where 
$\Delta V_M$ is of order $\sim$ 1 eV 
in the point-charge model with formal ionic charges, 
but is not detectable experimentally by photoemission spectroscopy 
within the accuracy of $\lesssim 100$ meV. 

In this study, we have investigated the effect of $\Delta V_M$ 
from the shifts of the core-level photoemission spectra of 
La$_{0.6}$Sr$_{0.4}$MnO$_3$ (LSMO) thin films 
under various epitaxial strain from 
three kinds of substrates, SrTiO$_3$ (STO), 
(LaAlO$_3$)$_{0.3}$-(SrAl$_{0.5}$Ta$_{0.5}$O$_3$)$_{0.7}$ (LSAT), 
and LaAlO$_3$ (LAO). Konishi {\it et al.} \cite{Konishi} 
have reported that the electronic structures of LSMO thin films 
are affected by strains from the substrates. It 
becomes a ferromagnetic metal on STO and LSAT and 
a $C$-type antiferromagnetic insulator on LAO. 
Among these three samples, one expects 
that $V_M$ (and $\mu$ to a lesser extent) changes 
but that the other terms in Eq.~(1) are kept unchanged 
because the band filling remains unchanged. 
We found that the amount of the shifts was of order 
$\sim$ 100 meV and is similar 
to that predicted 
by the Madelung potential, 
indicating that changes in $V_M$ are 
experimentally observed. In particular, 
the experimental shifts of the La $4d$ and Sr $3d$ core levels are 
almost the same as the calculation, whereas those of the 
O $1s$ and Mn $2p$ core levels are negligible, in disagreement 
with the calculation. We interpret these results in terms of 
the strong covalent character of the Mn-O bonding. 

\section{Experiment and calculation}
The LSMO thin films were fabricated in a laser MBE
chamber and transferred in vacuum to a photoemission
chamber at BL-2C of the Photon Factory \cite{Horiba}. 
The films were grown on the three substrates, 
LAO (001) (which causes compressive strain), 
LSAT (001) (almost no strain), 
and STO (001) (tensile strain). 
The thickness of the films were about 40 nm. 
We confirmed that all the films exhibited the same electrical 
and magnetic properties as those reported in the previous 
studies \cite{Konishi, HoribasubHX}. 
The lattice constants were determined by four-circle x-ray 
diffraction as summarized in Table~\ref{tab1}. 
The photoemission spectra were taken using a Gammadata Scienta 
SES-100 spectrometer. All the spectra were measured at room 
temperature. The total energy resolution was about 200 meV. 
The chemical potential 
was determined by measuring the spectra of gold which is 
in electrical contact with the thin films. The Madelung 
potentials were calculated 
by Ewald's method \cite{Kittelbook}. 
\begin{table}
\begin{center}
\caption{In-plane ($a$-axis) and out-of plane ($c$-axis) 
lattice constants of LSMO thin films grown on LAO, LSAT 
and STO substrates.}
\begin{tabular}{ccccc}
\hline
Substrate & $a$ ($\mbox{\AA}$) & $c$ ($\mbox{\AA}$) 
& $c/a$ & $V$ ($\mbox{\AA}^3$)\\
\hline
LAO & 3.79 & 3.98 & 1.05 & 57.2\\
LSAT & 3.87 & 3.87 & 1.00 & 58.0\\
STO & 3.91 & 3.83 & 0.98 & 58.6\\
\hline
\end{tabular}
\label{tab1}
\end{center}
\end{table}

\section{Results and Discussion}
\begin{figure}
\begin{center}
\includegraphics[width=9cm]{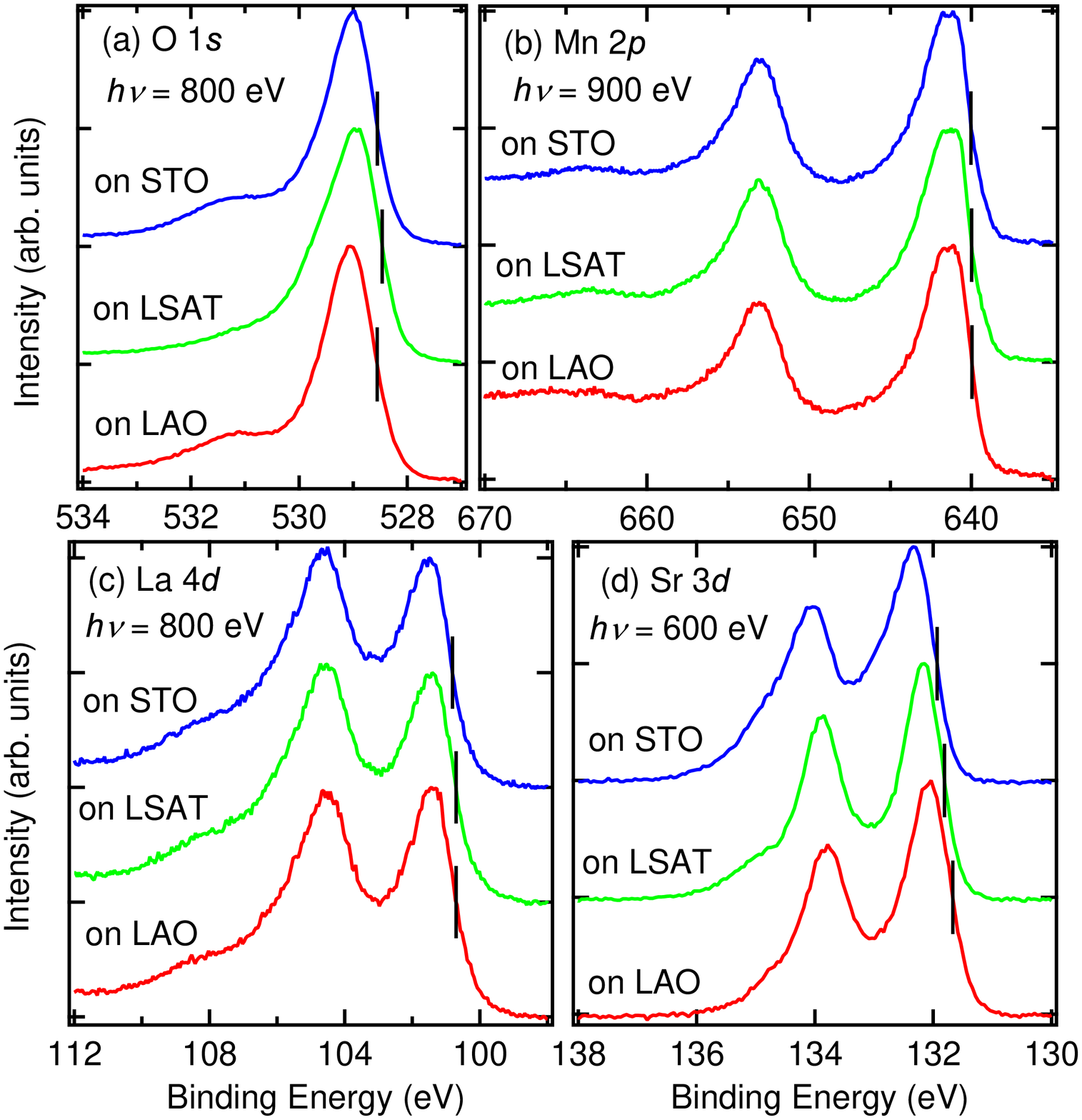}
\caption{(Color online) 
Core-level photoemission spectra of LSMO thin films epitaxially 
grown on STO, LSAT, and LAO substrates. 
(a) O $1s$, (b) Mn $2p$, (c) La $4d$, and (d) Sr $3d$.}
\label{LSMOcore}
\end{center}
\end{figure}

Figure \ref{LSMOcore} shows the core-level photoemission 
spectra of the LSMO thin films. 
The weakness of contamination signals on the higher-binding energy
side of the O $1s$ peak indicates that the surface was 
reasonably clean owing to the {\it in-situ} measurements. 
One notices here that the experimental shifts of 
the core levels are on the order of 
$\sim$ 100 meV. 

\begin{figure}
\begin{center}
\includegraphics[width=9cm]{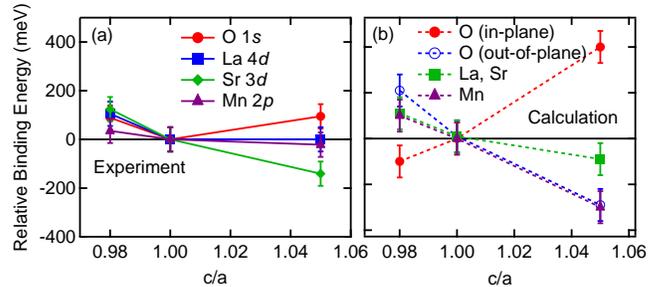}
\caption{(Color online) 
Binding-energy shifts of each core level in LSMO
thin films. (a) Experimental values determined by
photoemission spectroscopy. (b) Calculated values
determined from the effect of $\Delta V_M$ calculated 
using the point-charge model with formal ionic charges.}
\label{LSMOcore2}
\end{center}
\end{figure}

Figure \ref{LSMOcore2} shows the binding-energy shifts 
of each core level of the LSMO thin films obtained 
from experiment (photoemission spectroscopy) [panel (a)] 
and Madelung-potential calculation [panel (b)]. 
To calculate the Madelung potentials, 
we assumed formal ionic charges for all the atoms, 
that is, 
$3+$ for La, 
$2+$ for Sr, 
$2-$ for O and $3.4+$ for Mn. 
The energy scales of the shifts ($\sim$ 200 meV) 
are similar to experiment ($\sim$ 100 meV), 
which means that changes 
in $V_M$ are experimentally observed. 
If we look into more details, however, 
different behaviors are seen between experiment and calculation. 
For the O $1s$ core level, the calculated Madelung potentials felt 
by the in-plane and out-of-plane O atoms are quite different 
and should cause a splitting of the O $1s$ 
core-level photoemission spectra, 
but we do not observe such a splitting 
as shown in Fig.~\ref{LSMOcore} (a). 

\begin{figure}
\begin{center}
\includegraphics[width=9cm]{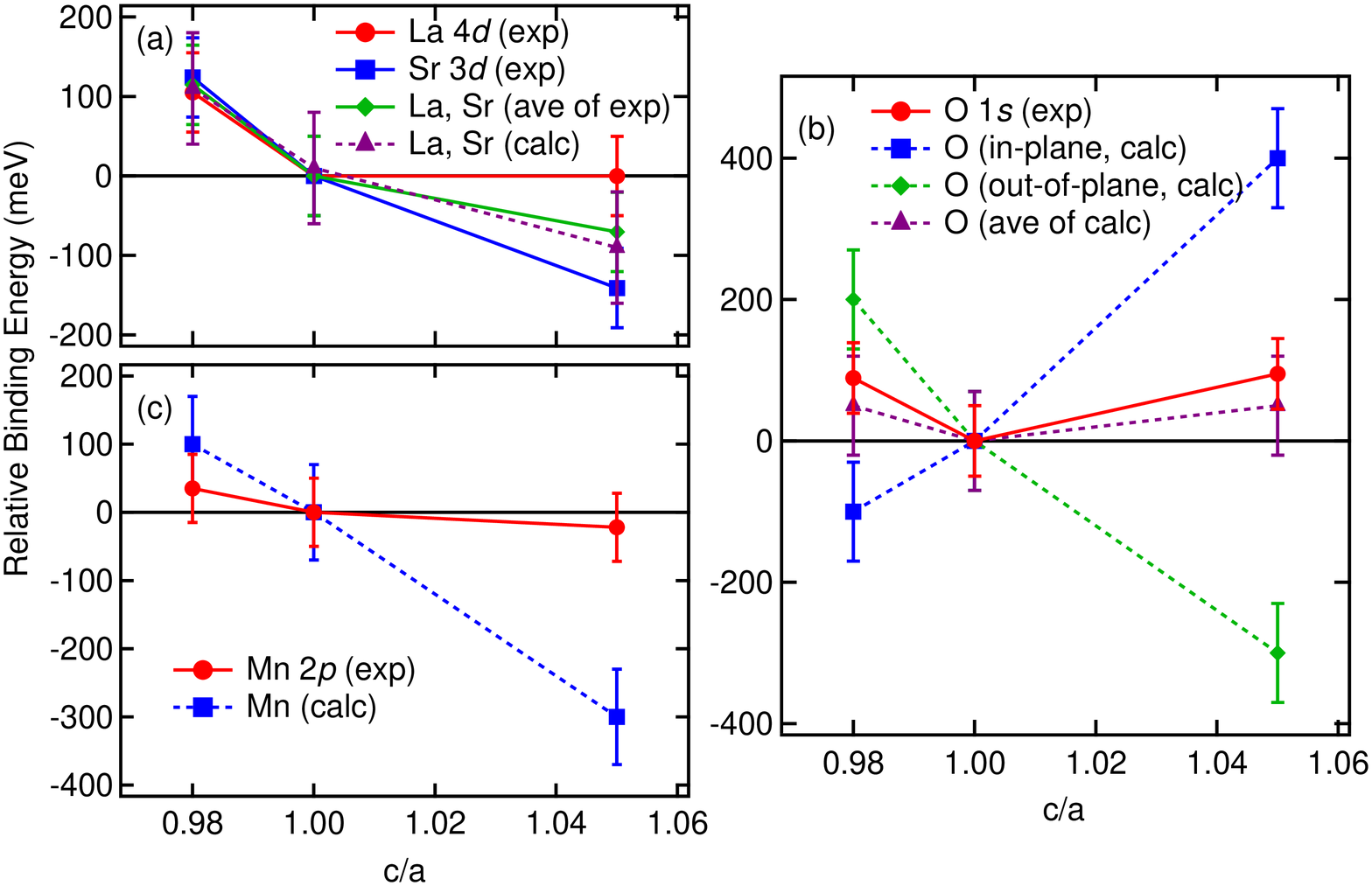}
\caption{(Color online) 
Comparison of the binding-energy shifts 
of each core level in LSMO thin films 
between experiment and calculation for the model of 
La$^{3+}_{0.6}$Sr$^{2+}_{0.4}$Mn$^{3.4+}$O$^{2-}_3$. 
(a) La $4d$ and Sr $3d$, (b) O $1s$, and (c) Mn $2p$.}
\label{LSMOcore3}
\end{center}
\end{figure}

To compare the experiment and calculation for each core level, 
we show such comparison in Fig.~\ref{LSMOcore3}. 
Interestingly, as shown in Fig.~\ref{LSMOcore3} (a), 
the experimental shifts of La $4d$ and Sr $3d$ 
are almost the same as the calculation. 
We consider that the effects of the changes in the 
Madelung potential on these core-levels are directly 
reflected on their binding energy shifts 
probably because the La and Sr 
ions do not formally have valence electrons 
that can suppress changes in the Madelung potential 
through covalency effect. 
Figure \ref{LSMOcore3} (b) and (c), 
on the other hand, shows that 
the experimental shifts of the O $1s$ and Mn $2p$ core levels 
are negligibly small compared with the calculation. 
This may be attributed to the strong covalency 
of the Mn-O bonding. 

\begin{figure}
\begin{center}
\includegraphics[width=9cm]{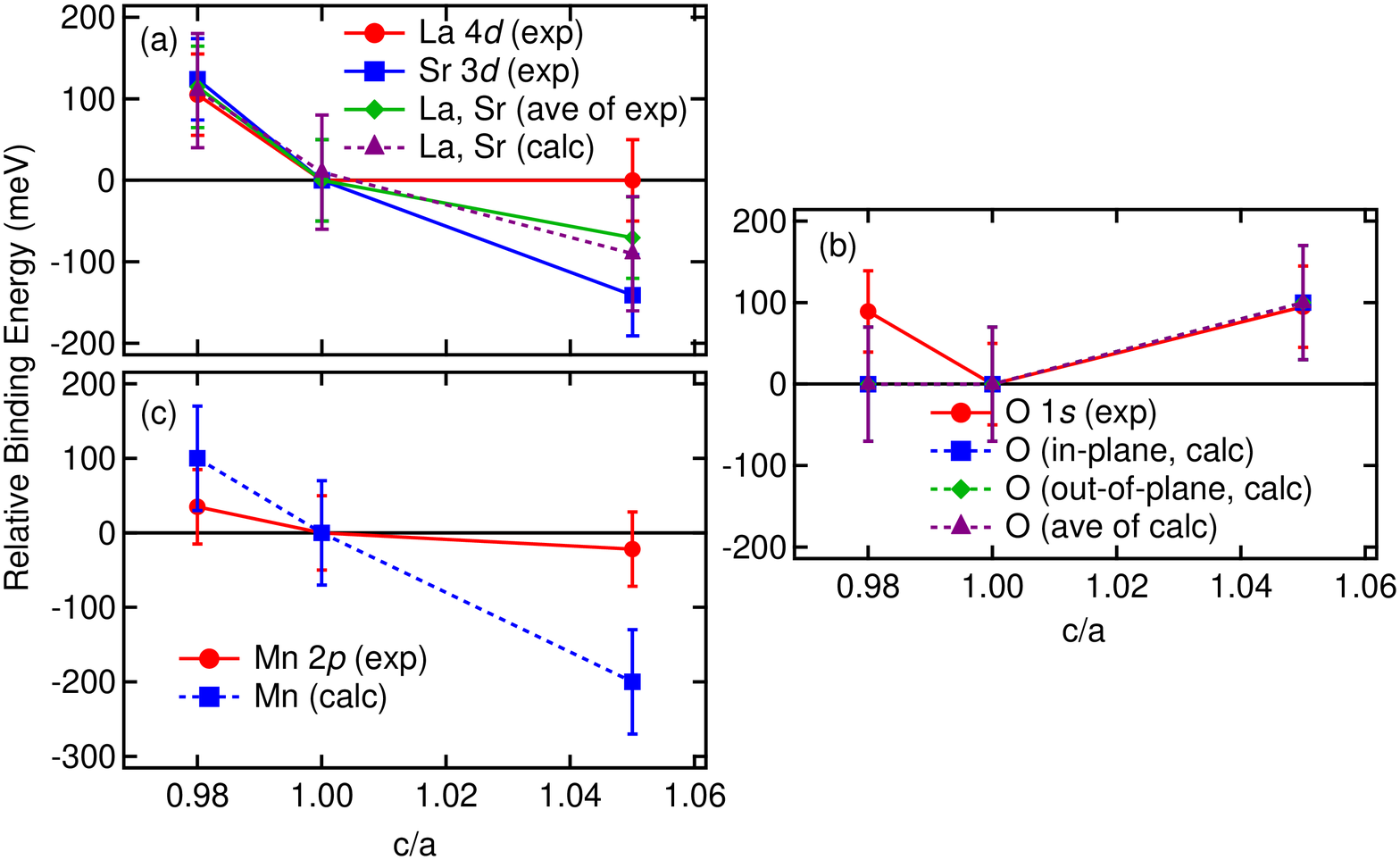}
\caption{(Color online) 
Comparison of the binding-energy shifts 
of each core level in LSMO thin films 
between experiment and calculation for the model of 
La$^{3+}_{0.6}$Sr$^{2+}_{0.4}$Mn$^{1.9+}$O$^{1.5-}_3$. 
(a) La $4d$ and Sr $3d$, (b) O $1s$, and (c) Mn $2p$.}
\label{LSMOcore4}
\end{center}
\end{figure}

\begin{figure}
\begin{center}
\includegraphics[width=9cm]{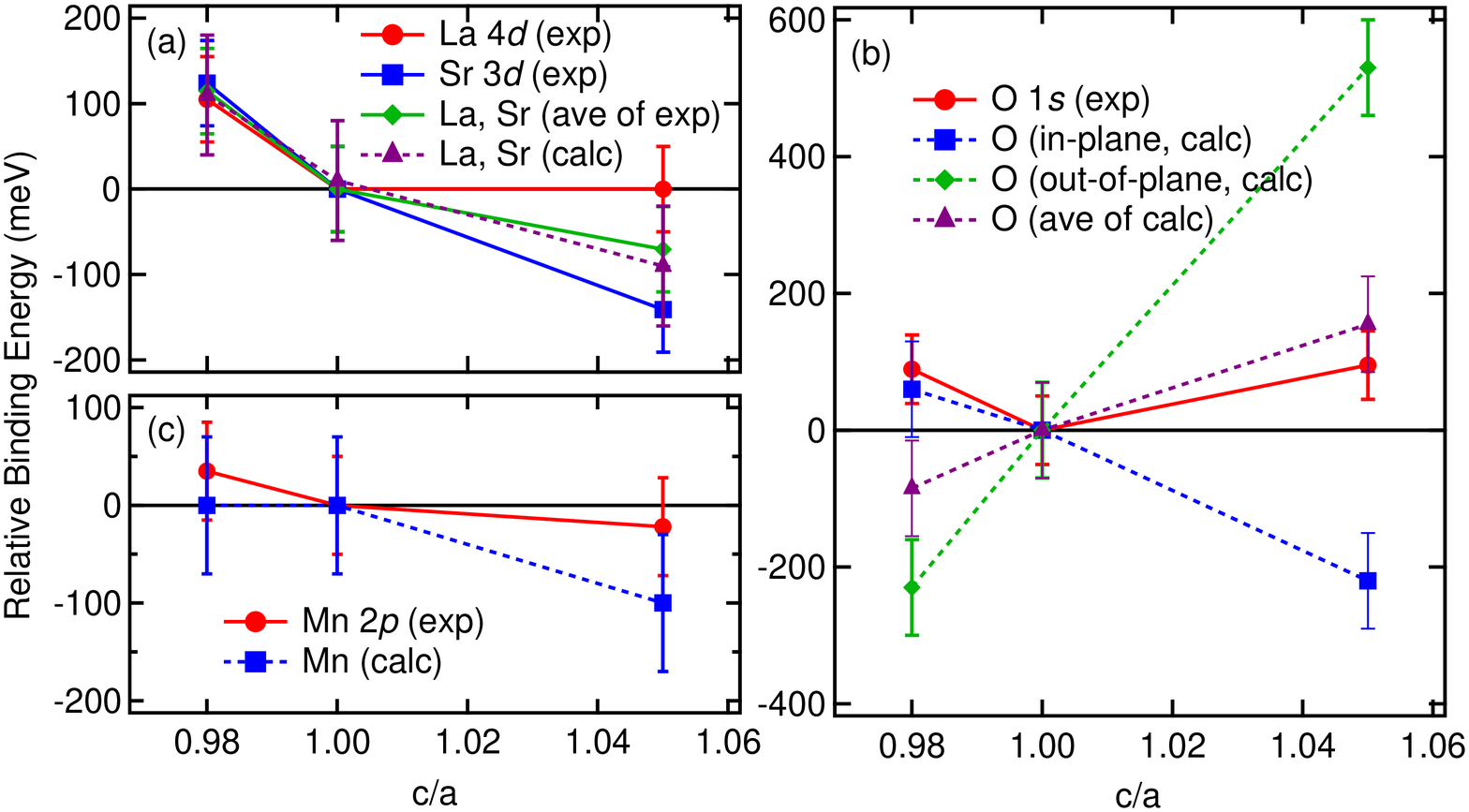}
\caption{(Color online) 
Comparison of the binding-energy shifts 
of each core level in LSMO thin films 
between experiment and calculation for the model of 
La$^{3+}_{0.6}$Sr$^{2+}_{0.4}$Mn$^{0.4+}$O$^{-}_3$. 
(a) La $4d$ and Sr $3d$, (b) O $1s$, and (c) Mn $2p$.}
\label{LSMOcore5}
\end{center}
\end{figure}

To simulate the effects of covalency between the Mn 
and O atoms, which may reduce the Madelung 
potential changes, we also calculated the Madelung 
potentials in two other models in which 
the charges on the O atoms are reduced 
from the ionic value of $2-$ and 
the charges on La and Sr atoms 
retain the ionic values. 
In the first model, 
O is assumed to be $1.5-$ 
and Mn is $1.9+$, and 
in the second model 
O is assumed to be $1-$ 
and Mn is $0.4+$. 
The calculated results 
are compared with experiment in 
Figs.~\ref{LSMOcore4} and \ref{LSMOcore5}, respectively. 
As for the La $4d$ and Sr $3d$ core levels, 
agreement between experiment and calculation remains good, 
which means that this agreement does not depend on 
the Mn-O covalency. 
The calculated O $1s$ core level does not split and shows 
good agreement with experiment for the 
$\mbox{O}^{1.5-}$ model as shown in Fig.~\ref{LSMOcore4} (b). 
As for Mn $2p$, the agreement becomes better as we decrease 
the charge on O atoms from $-2$ to $-1$, as shown in 
Figs.~\ref{LSMOcore4} (c) and \ref{LSMOcore5} (c). 
From the above results, one can see that 
better agreement between experiment and 
calculation is obtained when the effects of 
strong covalency of the Mn-O bonding are 
taken into account through the reduction of the 
electric charges on the Mn and O atoms. 

The present scenario described above 
is also consistent with the negligible effects of 
$\Delta V_M$ when the band filling is varied. 
In the case of filling control, the doped carriers 
would be relatively uniformly distributed 
between the Mn and O atoms 
and $\Delta V_M$ is kept almost 
unchanged in all the core levels \cite{Ino, pote}. 

In the present analysis, we have ignored changes in 
the chemical potential $\Delta \mu$ under 
the epitaxial strain, which would be small 
but in principle exists. 
The general increase of the binding energies in the 
strained samples (on the STO and LAO substrates) 
relative to the unstrained one (on the LSAT substrate) 
seen in experiment in Fig.~2 (a) 
suggests upward shifts of the chemical potential 
for non-zero strain $q (\equiv c/a-1)$. 
In fact, the lowest order shift with respect 
to $q$ is $\Delta \mu \propto q^2$ in cubic systems. 
More systematic 
studies on transition-metal-oxide thin films on 
various substrates will further clarify this effect. 

\section{Summary}
We have investigated the effect of Madelung potentials in the 
core-level photoemission spectra of LSMO thin films 
epitaxially grown on various substrates. The amount of binding-energy 
shifts ($\sim$ 100 - 200 meV) is similar in both experiment and 
calculation, indicating that changes in Madelung potentials 
are experimentally observed. 
As for the La $4d$ and Sr $3d$ core levels, 
the experimentally observed 
shifts were almost the same as the calculation. 
The experimental shifts of the O $1s$ 
and Mn $2p$ core levels were negligibly small, 
in disagreement with the calculation. 
These behaviors can be explained by 
the strong covalent character of the Mn-O bonding. 

\section*{Acknowledgment}
We are grateful to K. Ono and A. Yagishita for their support 
at KEK-PF. This work was supported by a Grant-in-Aid 
for Scientific Research (A19204037) from 
the Japan Society for the Promotion of 
Science (JSPS) and a Grant-in-Aid 
for Scientific Research in Priority Areas 
``Invention of Anomalous Quantum Materials'' 
from the Ministry of Education, Culture, 
Sports, Science and Technology. 
H.W. acknowledges financial support from JSPS. 
The work was done under the approval of the Photon Factory 
Program Advisory Committee (Proposal Nos.~2005G101 and 2005S2-002) 
at the Institute of Material Structure Science, KEK. 
\bibliography{LVO1tex}
\end{document}